\documentclass[a4paper,12pt,leqno]{article}
\usepackage{theorem}
\newtheorem{thm}{THEOREM}
\newtheorem{lem}[thm]{LEMMA}
\newtheorem{cor}[thm]{COROLLARY}

\theoremheaderfont{\scshape}

\newcommand{\ket}[1]{| #1 \rangle}

\title{{\Large {\bf ABSORPTION PROBLEMS FOR QUANTUM WALKS IN ONE DIMENSION}
}}
\author{{\small By NORIO KONNO, TAKAO NAMIKI$^{*}$, TAKAHIRO SOSHI AND AIDAN SUDBURY$^{**}$} \\
{\small
{\it Yokohama National University, $^{*}$Hokkaido University and $^{**}$Monash University}}
}

\date{\empty }
\begin{document}
\maketitle

\par\noindent
\begin{small}
{\bf Abstract}. 
This paper treats absorption problems for the one-dimensional quantum walk determined by a $2 \times 2$ unitary matrix $U$ on a state space $\{0,1, \ldots, N\}$ where $N$ is finite or infinite by using a new path integral approach based on an orthonormal basis $P, Q, R$ and $S$ of the vector space of complex $2 \times 2$ matrices. Our method studied here is a natural extension of the approach in the classical random walk.

\footnote[0]{
{\it Abbr. title:} Absorption problems of quantum walks.
}
%{\it AMS 2000 subject classifications.}
%60F05, 60G50, 82B41, 81Q99


\end{small}

\setcounter{equation}{0}
\section{Introduction}
\newcommand{\U}{\bar{U}} 
In this paper we consider absorption problems for quantum walks located on the sets $\{0,1, \ldots, N\}$ or $\{0,1, \ldots \}$. Before we move to a quantum case, first we describe the classical random walk on a finite set $\{0,1, \ldots, N\}$ with two absorbing barriers at locations $0$ and $N$ (see Grimmett and Stirzaker (1992), Durrett (1999), for examples). The particle moves at each step either one unit to the left with probability $p$ or one unit to the right with probability $q=1-p$ until it hits one of the absorbing barriers. The directions of different steps are independent of each other. The classical random walk starting from $k \in \{0,1, \ldots, N \}$ at time $n$ is denoted by $S^k _n$ here. Let 
\[
T_m = \min \{ n \ge 0 : S^k _n = m \}
\]
be the time of the first visit to $m \in \{0,1, \ldots, N \}$. Using the subscript $k$ to indicate $S^k _0=k$, we let
\[
P^{(N)} _k = P_k (T_0 < T_N)
\]
be the probability that the particle hits $0$ starting from $k$ before it arrives at $N$. The absorption problem is well known as the Gambler's ruin problem. 
\par
Recently, in quantum walks, various problems including absorption  have been widely investigated by a number of groups in connection with  quantum computing. Examples include Aharonov {\it et al.} (2001), Ambainis {\it et al.} (2001), Bach {\it et al.} (2002), Childs, Farhi and Gutmann (2002), D\"ur {\it et al.} (2002), Kempe (2002), Konno, Namiki and Soshi (2002), Konno (2002a, 2002b), Mackay {\it et al.} (2002), Moore and Russell (2001), Travaglione and Milburn (2002), Yamasaki, Kobayashi and Imai (2002). For a more general setting including quantum cellular automata, see Meyer (1996). A more mathematical point of view  can be found in  Brylinsky and Chen (2002).

\par
The study of absorption problems in the classical case is important in many fields including physics, biology, and computer science. The behavior of quantum walks differs from that of classical random walks in several striking ways. Let $X_n ^{\varphi}$ be the quantum walk on a line with no absorbing boundaries at time $n$ starting from initial qubit state $\varphi$. In the symmetric classical case the distribution of the scaled position $S^0 _n / \sqrt{n}$ converges to $N(0,1)$, the standard normal, as $(n \to \infty)$. However, a differently rescaled quantum walk $X^{\varphi}_n /n$ converges to a random variable $Z^{\varphi}$ where $Z^{\varphi}$ has a density $1 / \pi (1-x^2) \sqrt{1-2x^2}$ for $x \in (- \sqrt{2}/2, \sqrt{2}/2)$ (see Konno (2002a, 2002b)). As a corollary, it is shown that, whereas the standard deviation $\sigma_n$ at time $n$ in the classical case is $\sqrt{n}$, in the quantum case it $~\sqrt{(2- \sqrt{2})} \times n \> (\approx 0.5412 \times n) $ for large $n$. It is hoped that quantum walks may have many new applications as in the classical case. The main interest of our paper is in studying quantum walks with one or two absorbing boundaries. 
    
\par
This paper is organized as follows. The quantum walk considered here is determined by a $2 \times 2$ unitary matrix $U$. In Section 2, we give a definition of the quantum walk and introduce a new path integral approach based on an orthonormal basis $P, Q, R,$ and $S$ of the vector space of complex $2 \times 2$ matrices which is called the PQRS method here. Section 3 provides some known results on absorption problems for classical and quantum cases. Section 4 is devoted to our new results. We summarize this paper in Section 5, which mentions a direction for further study. 

\section{Definition and PQRS Method}

The time evolution of the one-dimensional quantum walk studied here is given by the following unitary matrix (see Nielsen and Chuang (2000), Brylinsky and Chen (2002), for examples):
\begin{eqnarray*}
U=
\left[
\begin{array}{cc}
a & b \\
c & d
\end{array}
\right]
\end{eqnarray*}
\par\noindent
where $a,b,c,d \in {\bf C}$ and ${\bf C}$ is the set of complex numbers. So we have 
\begin{eqnarray*}
&& |a|^2 + |b|^2 =|c|^2 + |d|^2 =1, \qquad a \overline{c} + b \overline{d}=0, \\
&& c= - \triangle \overline{b}, \qquad d= \triangle \overline{a}
\end{eqnarray*}
where $\overline{z}$ is a complex conjugate of $z \in {\bf C}$ and $\triangle = \det U = ad - bc.$ We should note that the unitarity of $U$ gives $|\triangle|=1.$
\par
The quantum walk is a quantum generalization of the classical random walk with an additional degree of freedom called the chirality. The chirality takes values left and right, and means the direction of the motion of the particle. The evolution of the quantum walk is given by the following way. At each time step, if the particle has the left chirality, it moves one unit to the left, and if it has the right chirality, it moves one unit to the right.

\par 
The unitary matrix $U$ acts on two chirality states $\ket{L}$ and $\ket{R}$:
\begin{eqnarray*}
&& \ket{L} \>\> \to \>\> a\ket{L} + c\ket{R} \\
&& \ket{R} \>\> \to \>\> b\ket{L} + d\ket{R} 
\end{eqnarray*}
where $L$ and $R$ refer to the right and left chirality state respectively. In fact, define
\begin{eqnarray*}
\ket{L} = 
\left[
\begin{array}{cc}
1 \\
0  
\end{array}
\right],
\qquad
\ket{R} = 
\left[
\begin{array}{cc}
0 \\
1  
\end{array}
\right]
\end{eqnarray*}
so we have
\begin{eqnarray*}
&& U\ket{L} = a\ket{L} + c\ket{R} \\
&& U\ket{R} = b\ket{L} + d\ket{R} 
\end{eqnarray*}
\par
More precisely, at any time $n$, the amplitude of the location of the particle is defined by a 2-vector $\in {\bf C}^2$ at each location $\{0,1,...,N\}$. The probability the particle is at location $j$ is given by the square of the modulus of the vector at $j$. If $\ket{\Psi_j (n)}$ define the amplitude at time $n$ at location $j$ where
\begin{eqnarray*}
\ket{\Psi_j (n)} = \left[
\begin{array}{cc}
\psi_j ^L (n) \\
\psi_j ^R (n)
\end{array}
\right]
\end{eqnarray*}
with the chirality being left (upper component) or right (lower component), then the dynamics for $\ket{\Psi_j (n)}$ in quantum walks is given by the following transformation:
\begin{eqnarray*}
\ket{\Psi_j (n+1)} = \ket{L} \langle L | U \ket{\Psi_{j+1} (n)}
+ \ket{R} \langle R | U \ket{\Psi_{j-1} (n)}
\end{eqnarray*}
The above equation can be rewritten as
\begin{eqnarray}
\ket{\Psi_j (n+1)} = P \ket{\Psi_{j+1} (n)} + Q \ket{\Psi_{j-1} (n)}
\end{eqnarray}
where
\begin{eqnarray*}
P= 
\left[
\begin{array}{cc}
a & b \\
0 & 0 
\end{array}
\right], 
\quad
Q=
\left[
\begin{array}{cc}
0 & 0 \\
c & d 
\end{array}
\right]
\end{eqnarray*}
We see $U=P+Q.$ The unitarity of $U$ ensures that the amplitude always defines a probability distribution for the location.

\par
The simplest and well-studied example of a quantum walk is the Hadamard walk whose unitary matrix $U$ is defined by 
\begin{eqnarray*}
H = 
{1 \over \sqrt{2}}
\left[
\begin{array}{cc}
1 & 1 \\
1 & -1 
\end{array}
\right] 
\end{eqnarray*}
The dynamics of this walk corresponds to that of the symmetric random walk in the classical case. In general, the following unitary matrices can also lead to symmetric walks:
\begin{eqnarray*}
U_{\eta, \phi, \psi} = {e^{i \eta} \over \sqrt{2}}
\left[
\begin{array}{cc}
e^{i(\phi + \psi)} & e^{-i(\phi - \psi)}  \\
e^{i(\phi - \psi)}  & -e^{-i(\phi + \psi)}  
\end{array}
\right] 
\end{eqnarray*}
where $\eta, \phi ,$ and $\psi$ are real numbers (see pp.175-176 in Nielsen and Chuang (2000), for example). In particular, we see $U_{0,0,0}=H$.  
\par
However symmetry of the Hadamard walk depends heavily on the initial qubit state, see Konno, Namiki and Soshi (2002). Another extension of the Hadamard walk is:
\begin{eqnarray*}
H (\rho) = 
\left[
\begin{array}{cc}
\sqrt{\rho} & \sqrt{1-\rho} \\
\sqrt{1-\rho} & - \sqrt{\rho} 
\end{array}
\right] 
\end{eqnarray*}
where $0 \le \rho \le 1$. Note that  $\rho = 1/2$ is the Hadamard walk, that is, $H=H(1/2)$.

In the present paper, the study on the dependence of some important quantities (e.g., absorption probability) on initial qubit state is one of the essential parts, so we define the collection of initial qubit states as follows:
\[
\Phi = \left\{ \varphi =
\left[
\begin{array}{cc}
\alpha \\
\beta   
\end{array}
\right]
\in 
{\bf C}^2
:
|\alpha|^2 + |\beta|^2 =1
\right\}
\]

Let $X_n ^{\varphi}$ be the quantum walk at time $n$ starting from initial qubit state $\varphi \in \Phi$ with $X_0 ^{\varphi}=0.$ In contrast with classical random walks, $X_n ^{\varphi}$ can not be written as $X_n ^{\varphi} = Y_1 + \cdots + Y_n$ where $Y_1, Y_2, \ldots $ are independent and identically distributed random variables. 
In our treatment of quantum walks, as well as the matrices $P$ and $Q$, it is convenient to introduce
\[
R=
\left[
\begin{array}{cc}
c & d \\
0 & 0 
\end{array}
\right], 
\quad
S=
\left[
\begin{array}{cc}
0 & 0 \\
a & b 
\end{array}
\right]
\]
We should remark that $P,Q,R,$ and $S$ form an orthonormal basis of the vector space of complex $2 \times 2$ matrices $M_2 ({\bf C})$ with respect to the trace inner product $\langle A|B \rangle = \>$ tr $(A^{\ast}B)$. Therefore we can express any $2 \times 2$ matrix $A$ conveniently in the form,
\begin{eqnarray} 
A = 
\hbox{tr} (P^{\ast} A)P + \hbox{tr} (Q^{\ast} A)Q + \hbox{tr} (R^{\ast} A)R + 
\hbox{tr} (S^{\ast} A)S 
\end{eqnarray} 
We call the analysis based on $P,Q,R,$ and $S$ the PQRS method. 

The $n \times n$ unit and zero matrices are written $I_n$ and $O_n$ respectively. For instance, if $A = I_2$, then 
\begin{eqnarray} 
I_2 = \overline{a} P + \overline{d} Q + \overline{c} R + \overline{b} S
\end{eqnarray} 
The next table of products of $P,Q,R,$ and $S$ is very useful in computing some quantities:
\par
\
\par
\begin{center}
\begin{tabular}{c|cccc}
  & $P$ & $Q$ & $R$ & $S$  \\ \hline
$P$ & $aP$ & $bR$ & $aR$ & $bP$  \\
$Q$ & $cS$ & $dQ$& $cQ$ & $dS$ \\
$R$ & $cP$ & $dR$& $cR$ & $dP$ \\
$S$ & $aS$ & $bQ$ & $aQ$ & $bS$ 
\end{tabular}
\end{center}
where $PQ=bR$, for example. 
\par
Now we describe the evolution and measurement of quantum walks starting from location $k$ on $\{ 0, 1, \ldots , N \}$ with absorbing boundaries (see Ambainis {\it et al.} (2001), Bach {\it et al.} (2002), and Kempe (2002) for more detailed information, for examples). 
\par
First we consider $N= \infty$ case. In this case, an absorbing boundary is present at location 0. The evolution mechanism is described as follows:
\par 
Step 1. Initialize the system $\varphi \in \Phi$ at location $k$.
\par
Step 2. (a) Apply Eq. (2.1) to one step time evolution. (b) Measure the system to see where it is or is not at location 0.
\par 
Step 3. If the result of measurement revealed that the system was at location 0, then terminate the process, otherwise repeat step 2.
\par
In this setting, let $\Xi^{(\infty)} _k (n)$ 
be the sum over possible paths for which the particle first hits 0 at time $n$ starting from $k$. For example, 
\begin{eqnarray} 
\Xi^{(\infty)} _1 (5) = P^2 Q P Q + P^3 Q^2= (a b^2 c + a^2bd) R
\end{eqnarray} 
The probability that the particle first hits 0 at time $n$ starting from $k$ is 
\begin{eqnarray} 
P^{(\infty)} _k (n; \varphi) = | \Xi^{(\infty)} _k (n) \varphi |^2
\end{eqnarray} 
So the probability that the particle first hits 0 starting from $k$ is 
\begin{eqnarray*} 
P^{(\infty)} _k (\varphi) = \sum_{n=0} ^{\infty} P^{(\infty)} _k (n; \varphi) 
\end{eqnarray*} 
\par
Next we consider $N < \infty$ case. This case is similar to the $N= \infty$ case, except that two absorbing boundaries are presented at locations 0 and $N$ as follows:
\par 
Step 1. Initialize the system $\varphi \in \Phi$ at location $k$.
\par
Step 2. (a) Apply Eq. (2.1) to one step time evolution. (b) Measure the system to see where it is or is not at location 0. (c) Measure the system to see where it is or is not at location $N$.
\par 
Step 3. If the result of either measurement revealed that the system was either at location 0 or location $N$, then terminate the process, otherwise repeat step 2.
\par
Let $\Xi^{(N)} _k (n)$ be the sum over possible paths for which the particle first hits 0 at time $n$ starting from $k$ before it arrives at $N$. For example,
\begin{eqnarray*} 
\Xi^{(N)} _1 (5) = P^2 Q P Q =a b^2 c R
\end{eqnarray*} 
In a similar way, we can define $P^{(N)} _k (n; \varphi)$ and $P^{(N)} _k (\varphi)$.

\section{Facts for the Classical and Quantum Cases}

In this section we review some results and conjectures on absorption problems related to this paper for both classical and quantum walks in one dimension. 

First we review the classical case. As we described in the Introduction, $P^{(N)} _k = P_k (T_0 < T_N)$ denotes the probability that the particle hits $0$ starting from $k$ before it arrives at $N$. We may use conditional probabilities to see that $P^{(N)} _k$ satisfies the following difference equation:
\begin{eqnarray}
P^{(N)} _k = p P^{(N)} _{k-1} + q P^{(N)} _{k+1} \qquad (1 \le k \le N-1)
\end{eqnarray}
with boundary conditions:
\begin{eqnarray}
P^{(N)} _0=1, \quad  P^{(N)} _N=0
\end{eqnarray}
The solution of such a difference equation is given by
\begin{eqnarray}
&& P^{(N)} _k = 1  - {k \over N} \quad \hbox{if} \qquad p = 1/2 \\
&& P^{(N)} _k = {(p/q)^k - (p/q)^N \over 1 - (p/q)^N} \qquad \hbox{if} \quad p \not= 1/2
\end{eqnarray}
for any $0 \le k \le N$. Therefore , when $N=\infty$, we see that
\begin{eqnarray}
&& P^{(\infty)} _k = 1 \quad \hbox{if} \qquad  1/2 \le p \le 1 \\
&& P^{(\infty)} _k = (p/q)^k \qquad \hbox{if} \quad 0 \le p < 1/2
\end{eqnarray}
Furthermore, 
\begin{eqnarray}
&& \lim_{k \to \infty} P^{(\infty)} _k = 1 \qquad \hbox{if} \qquad 1/2 \le p \le 1\\
&& \lim_{k \to \infty} P^{(\infty)} _k = 0 \qquad \hbox{if} \quad 0 \le p < 1/2
\end{eqnarray}

Let $T_0$ be the first hitting time to 0. We consider the conditional expectation of $T_0$ starting from $k=1$ given the event $\{T_0 < \infty \}$, that is, $E_1 ^{(\infty)}(T_0|T_0 < \infty)= E_1 ^{(\infty)} (T_0 ; T_0 < \infty)/P_1 ^{(\infty)} ( T_0 < \infty) = E_1 ^{(\infty)} (T_0 ; T_0 < \infty)/P_1 ^{(\infty)}$
\begin{eqnarray*}
&& E_1 ^{(\infty)}(T_0|T_0 < \infty) =  {2p \over \sqrt{1-4pq}} - 1 
\qquad \hbox{if} \quad  1/2 < p \le 1 \\
&& E_1 ^{(\infty)}(T_0|T_0 < \infty) = \infty 
\qquad \hbox{if} \quad   p=1/2 \\
&& E_1 ^{(\infty)}(T_0|T_0 < \infty) =  {2q \over \sqrt{1-4pq}} - 1 
\qquad \hbox{if} \quad  0 \le p < 1/2  
\end{eqnarray*}

Next we review the quantum case. In the case of $U=H$ (the Hadamard walk), when $N= \infty$, that is, when the state space is $\{0,1, \ldots, \}$ case, Ambainis {\it et al.} (2001) proved
\begin{eqnarray}
P^{(\infty)} _1 ({}^t[0,1]) = P^{(\infty)} _1 ({}^t[1,0]) = {2 \over \pi}
\end{eqnarray}
and Bach {\it et al.} (2002) showed 
\begin{eqnarray*} 
\lim_{k \to \infty} P^{(\infty)} _k (\varphi) 
=  \left({1 \over 2} \right) |\alpha|^2 
+ \left( {2 \over \pi} - {1 \over 2} \right)|\beta|^2
+ 2 \left( {1 \over \pi} - {1 \over 2} \right) \Re(\overline{\alpha}\beta)
\end{eqnarray*}
for any initial qubit state $\varphi ={}^t [\alpha, \beta] \in \Phi.$ Furthermore, in the case of $U=H(\rho)$, Bach {\it et al.} (2002) gave
\begin{eqnarray*} 
&& \lim_{k \to \infty} P^{(\infty)} _k ({}^t[0,1]) 
= {\rho \over 1- \rho} \left( {\cos^{-1} (1- 2\rho) \over \pi} -1 \right) 
+ { 2 \over \pi \sqrt{1/\rho -1}} \\
&& \lim_{k \to \infty} P^{(\infty)} _k ({}^t[1,0]) 
= {\cos^{-1} (1- 2\rho) \over \pi} 
\end{eqnarray*}  
The second result was conjectured by Yamasaki, Kobayashi and Imai (2002).

When $N$ is finite the following conjecture by Ambainis {\it et al.} (2001) is still open for the $U=H$ case:
\begin{eqnarray*} 
P^{(N+1)} _1 ({}^t[0,1]) ={ 2 P^{(N)} _1 ({}^t[0,1]) +1 \over 2 P^{(N)} _1 ({}^t[0,1]) +2} \quad (N \ge 1), \qquad P^{(1)} _1 ({}^t[0,1]) =0. 
\end{eqnarray*} 
Solving the above recurrence gives
\begin{eqnarray} 
P^{(N)} _1 ({}^t[0,1]) = {1 \over \sqrt{2}} \times {(3+2 \sqrt{2})^{N-1} -1 \over (3+2 \sqrt{2})^{N-1} +1} \quad (N \ge 1)
\end{eqnarray}
However in contrast with $P^{(\infty)} _1 ({}^t[0,1]) =2/\pi$, Ambainis {\it et al.} (2001) proved
\[ 
\lim_{N \to \infty} P^{(N)} _1 ({}^t[0,1])=1/\sqrt{2}
\]
It should be noted that $P^{(\infty)} _k = \lim_{N \to \infty} P^{(N)}_k$ for any $0 \le k \le N$ in the classical case (see Eqs. (3.8) - (3.11)).

We are not aware of results concerning  $E_1 ^{(\infty)}(T_0|T_0 < \infty)$ having been published, however we will give such a result in the next section.

\section{Our Results}

In the first half of this section, we consider a general setting including a hitting time to $N$ before it arrives at $0$ or a hitting time to $0$ before it arrives at $N$, when $N$ is finite. In the case of $N= \infty$, a similar argument holds, so we will omit it here.

Noting that $\{ P,Q,R,S \}$ is a basis of $M_2 ({\bf C})$, $\Xi^{(N)} _k (n)$ can be written as
\begin{eqnarray*} 
\Xi^{(N)} _k (n) = p^{(N)} _k (n) P + q^{(N)} _k (n) Q + r^{(N)} _k (n) R + s^{(N)} _k (n) S 
\end{eqnarray*} 
Therefore Eq. (2.5) implies 
\begin{eqnarray*} 
P^{(N)} _k (n; \varphi) = |\Xi^{(N)} _k (n) \varphi|^2
= C_1 (n) |\alpha|^2 + C_2 (n) |\beta|^2 + 
2 \Re(C_3 (n) \overline{\alpha} \beta)
\end{eqnarray*} 
where $\Re (z)$ is the real part of $z \in {\bf C}$, $\varphi = {}^t[\alpha, \beta] \in \Phi$ and 
\begin{eqnarray*}
&& C_1 (n)= |a p^{(N)} _k (n) + c r^{(N)} _k (n)|^2 + |a s^{(N)} _k (n) + c q^{(N)} _k (n)|^2 \\
&& C_2 (n) = |b p^{(N)} _k (n) + d r^{(N)} _k (n)|^2 + |b s^{(N)} _k (n) + d q^{(N)} _k (n)|^2 \\
&& C_3 (n) = 
\overline{(a p^{(N)} _k (n) + c r^{(N)} _k (n))} (b p^{(N)} _k (n) + d r^{(N)} _k (n)) \\  
&& \qquad \qquad \qquad + \overline{(a s^{(N)} _k (n) + c q^{(N)} _k (n))} (b s^{(N)} _k (n) + d q^{(N)} _k (n)) 
\end{eqnarray*}

From now on we assume $N \ge 3$. Noting that the definition of $\Xi^{(N)} _k (n)$, we see that for $1 \le k \le N-1$, 
\begin{eqnarray*} 
\Xi^{(N)} _k (n) = \Xi^{(N)} _{k-1} (n-1)P + \Xi^{(N)} _{k+1} (n-1)Q  
\end{eqnarray*} 
The above equation is a quantum version of the difference equation, i.e., Eq. (3.6) for the classical random walk. As an example, see Eq. (2.4). Then we have
\begin{eqnarray*} 
&& p^{(N)} _k (n) = a p^{(N)} _{k-1} (n-1)+ c r^{(N)} _{k-1} (n-1)\\ 
&& q^{(N)} _k (n) = d q^{(N)} _{k+1} (n-1)+ b s^{(N)} _{k+1} (n-1)\\ 
&& r^{(N)} _k (n) = b p^{(N)} _{k+1} (n-1)+ d r^{(N)} _{k+1} (n-1)\\ 
&& s^{(N)} _k (n) = c q^{(N)} _{k-1} (n-1)+ a s^{(N)} _{k-1} (n-1)
\end{eqnarray*} 

Next we consider boundary condition related to Eq. (3.7) in the classical case. When $k=N$, 
\[
P^{(N)} _N (0;\varphi) = | \Xi^{(N)} _N (0) \varphi |^2 =0
\]
for any $\varphi \in \Phi$. So we take $\Xi^{(N)} _N (0) = O_2$, i.e.,   
\[
p^{(N)} _N (0) = q^{(N)} _N (0) = r^{(N)} _N (0) = s^{(N)} _N (0) = 0
\]
If $k=0$, then
\[
P^{(N)} _0 (0;\varphi) = | \Xi^{(N)} _0 (0) \varphi |^2 =1
\]
for any $\varphi \in \Phi$. So we choose $\Xi^{(N)} _N (0) = I_2$. From Eq. (2.3), we have
\[
p^{(N)} _0 (0) = \overline{a}, \>\> q^{(N)} _0 (0) = \overline{d}, \>\> r^{(N)} _0 (0) = \overline{c}, \>\> s^{(N)} _0 (0) = \overline{ b} \>\>
\]
Let
\begin{eqnarray*} 
v^{(N)} _k (n) = 
\left[
\begin{array}{cc}
p^{(N)} _k (n) \\
r^{(N)} _k (n)   
\end{array}
\right] 
, \quad
w^{(N)} _k (n) = 
\left[
\begin{array}{cc}
q^{(N)} _k (n) \\
s^{(N)} _k (n)   
\end{array}
\right] 
\end{eqnarray*}
Then we see that for $n \ge 1$ and $1 \le k \le N-1$, 
\begin{eqnarray} 
&& v^{(N)} _k (n) = 
\left[
\begin{array}{cc}
a & c \\
0 & 0 
\end{array}
\right] 
v^{(N)} _{k-1} (n-1)
+
\left[
\begin{array}{cc}
0 & 0 \\
b & d 
\end{array}
\right] 
v^{(N)} _{k+1} (n-1)
\\
&& w^{(N)} _k (n) = 
\left[
\begin{array}{cc}
0 & 0 \\
c & a 
\end{array}
\right] 
w^{(N)} _{k-1} (n-1)
+
\left[
\begin{array}{cc}
d & b \\
0 & 0 
\end{array}
\right] 
w^{(N)} _{k+1} (n-1)
\quad
\end{eqnarray}
and for $1 \le k \le N$, 
\begin{eqnarray*}
&& v^{(N)} _0 (0) 
= 
\left[
\begin{array}{cc}
\overline{a}  \\
\overline{c}
\end{array}
\right] 
, \quad
v^{(N)} _k (0) 
= 
\left[
\begin{array}{cc}
0  \\
0
\end{array}
\right] 
\\
&& w^{(N)} _0 (0) 
= 
\left[
\begin{array}{cc}
\overline{d}  \\
\overline{b}
\end{array}
\right] 
, \quad
w^{(N)} _k (0) 
= 
\left[
\begin{array}{cc}
0  \\
0
\end{array}
\right] 
\end{eqnarray*}
Moreover,
\begin{eqnarray*}
v^{(N)} _0 (n) 
=
v^{(N)} _N (n) 
= 
w^{(N)} _0 (n) 
=
w^{(N)} _N (n) 
= 
\left[
\begin{array}{cc}
0  \\
0
\end{array}
\right] 
\quad
(n \ge 1)
\end{eqnarray*}

From now on we focus on $1 \le k \le N-1$ case. So we consider only $n \ge 1$ case. Moreover from the definition of $\Xi^{(N)} _k (n)$, it is easily shown that there exist only two types of paths, that is, $P \ldots P$ and $P \ldots Q$. Therefore we see that $q^{(N)} _k (n)=s^{(N)} _k (n)=0 \> (n \ge 1)$. So we have

\begin{lem}
\label{lem:lem1}
\begin{eqnarray*} 
&& P^{(N)} _k (\varphi) = \sum_{n=1} ^{\infty} P^{(N)} _k (n; \varphi) \\
&& P^{(N)} _k (n; \varphi) 
= C_1 (n) |\alpha|^2 + C_2 (n) |\beta|^2 + 2 \Re(C_3 (n) \overline{\alpha} \beta)
\end{eqnarray*} 
where $\varphi = {}^t[\alpha, \beta] \in \Phi$ and 
\begin{eqnarray*}
&& C_1 (n) = |a p^{(N)} _k (n) + c r^{(N)} _k (n)|^2 \\
&& C_2 (n) = |b p^{(N)} _k (n) + d r^{(N)} _k (n)|^2  \\
&& C_3 (n) = 
\overline{(a p^{(N)} _k (n) + c r^{(N)} _k (n))} (b p^{(N)} _k (n) + d r^{(N)} _k (n))  
\end{eqnarray*}
\end{lem}

To solve $P^{(N)} _k (\varphi)$, we introduce generating functions of $p^{(N)} _k (n)$ and $r^{(N)} _k (n)$ as follows:
\begin{eqnarray*}
&& p^{(N)} _k (z) = \sum_{n=1} ^{\infty} p^{(N)} _k (n) z^n \\
&& r^{(N)} _k (z) = \sum_{n=1} ^{\infty} r^{(N)} _k (n) z^n
\end{eqnarray*}
%Furthermore 
%\begin{eqnarray*}
%v^{(N)} _k (z) = \left[
%\begin{array}{cc}
%p^{(N)} _k (z)  \\
%r^{(N)} _k (z)
%\end{array}
%\right] 
%\end{eqnarray*}
By Eq. (4.16), we have
\begin{eqnarray*} 
&& p^{(N)} _k (z) = a z p^{(N)} _{k-1} (z) + cz r^{(N)} _{k-1} (z) \\
&& r^{(N)} _k (z) = b z p^{(N)} _{k+1} (z) + dz r^{(N)} _{k+1} (z) 
\end{eqnarray*} 
Solving these, we see that both $p^{(N)} _k (z)$ and $r^{(N)} _k (z)$ satisfy the same recurrence:   
\begin{eqnarray*} 
&& d p^{(N)} _{k+2} (z) - \left( \triangle z+{1 \over z} \right) p^{(N)} _{k+1} (z) + a p^{(N)} _{k} (z) = 0\\
&& d r^{(N)} _{k+2} (z) - \left( \triangle z+{1 \over z} \right) r^{(N)} _{k+1} (z) + a r^{(N)} _{k} (z) = 0
\end{eqnarray*} 
From the characteristic equations with respect to the above recurrences, we have the same roots: if $a \not= 0$, then
\begin{eqnarray*} 
\lambda_{\pm} = {\triangle z^2 + 1 \mp \sqrt{\triangle^2 z^4 + 2 \triangle ( 1- 2 |a|^2)z^2 + 1} \over 2 \triangle \overline{a} z} 
\end{eqnarray*} 
where $\triangle = \det U = ad - bc$.

From now on we consider mainly $U=H$ (the Hadamard walk) case with $N=\infty$. Remark that the definition of $\Xi_1 ^{(\infty)} (n)$ gives $p_1 ^{(\infty)}(n)=0 \> (n \ge 2)$ and $p_1 ^{(\infty)}(1)=1$. So we have $p^{(\infty)} _1 (z) = z$. Moreover noting $\lim_{k \to \infty} p^{(\infty)}  _k(z) < \infty$, the following explicit form is obtained:
\begin{eqnarray*}
     && p^{(\infty)} _k (z) = z \lambda_ +^{k-1}\\
     && r^{(\infty)} _k (z) = \frac{-1+\sqrt{z^4+1}}{z} \lambda_+^{k-1}
\end{eqnarray*}
where
\begin{eqnarray*} 
\lambda_\pm=\frac{z^2-1\pm\sqrt{z^4+1}}{\sqrt{2}z}.
\end{eqnarray*}
Therefore for $k=1$,
\begin{eqnarray*}
r^{(\infty)} _1 (z)=\frac{-1+\sqrt{z^4+1}}{z}
\end{eqnarray*} 
From the above equation and the definition of the hypergeometric series ${}_2F_1(a, b; c ;z)$, we have
\[
\sum_{n=1} ^{\infty} (r^{(\infty)} _1 (n)) ^2 z^n
=\sum_{n=1}^\infty \left(\begin{array}{c}1/2\\n\end{array}\right)^2 z^{4n-1}={{}_2F_1(-1/2,-1/2;1;z^4)-1 \over z}
\]  
On the other hand, it should be noted that
\[
{}_2F_1(a, b; c ;1)= 
{\Gamma (c) \Gamma (c-a-b) \over \Gamma (c-a) \Gamma (c-b)}
\qquad (\Re (a+b-c) < 0)
\]
where $\Gamma (z)$ is the gamma function defined by
\[
\Gamma (z) = \int_0 ^{\infty} e^{-t} t^{z-1} \> dt \qquad (\Re (z) >0)
\]
Therefore 
\begin{eqnarray*}
\sum_{n=1} ^{\infty} (r^{(\infty)} _1 (n)) ^2 = {\Gamma (1) \Gamma (2) \over \Gamma (3/2)^2 } -1 = {4 \over \pi} - 1
\end{eqnarray*}
since $\Gamma (1) = \Gamma (2) =1$ and $\Gamma (3/2)= \sqrt{\pi}/2.$ By Lemma 1
\begin{eqnarray*} 
&&P^{(\infty)} _1 (\varphi)= 
\sum_{n=1} ^{\infty} 
\Bigg[ 
{1 \over 2} \left\{ (p^{(\infty)} _1 (n)) ^2 + (r^{(\infty)} _1 (n)) ^2 \right\} \\
&& \qquad \quad 
+ p^{(\infty)} _1 (n) r^{(\infty)} _1 (n) (|\alpha|^2-|\beta|^2)
+{1 \over 2} \left\{ (p^{(\infty)} _1 (n)) ^2 - (r^{(\infty)} _1 (n)) ^2 \right\} (\alpha{\bar\beta}+{\bar\alpha}\beta)
\Bigg]
\end{eqnarray*} 
Note that $p^{(\infty)} _1 (n) r^{(\infty)} _1 (n)=0 \>\> (n \ge 1)$, since $p_1 ^{(\infty)}(n)=0 \> (n \ge 2), \>\> p_1 ^{(\infty)}(1)=1$ and 
$r^{(\infty)} _1 (1)=0$. So we have 
\begin{eqnarray}  
P^{(\infty)} _1 (\varphi) = {2 \over \pi} + 
2 \left(1 - {2 \over \pi} \right) \Re(\overline{\alpha} \beta)
\end{eqnarray} 
for any initial qubit state $\varphi = {}^t [\alpha, \beta] \in \Phi$. This result is a generalization of Eq. (3.14) given by Ambainis {\it et al.} (2001). From Eq. (4.18), we get the range of $P_1^{(\infty)} (\varphi)$:
\[
  \frac{4-\pi}{\pi} \le P_1^{(\infty)}(\varphi)\le 1
\]
the equality holds in the following cases:
\[
P_1^{(\infty)} (\varphi) =1 \mbox{ iff }
\varphi=\frac{e^{i\theta}}{\sqrt{2}}\left[\begin{array}{c}1\\1\end{array}\right]
\mbox{ and }
  P_1^{(\infty)} (\varphi) =\frac{4-\pi}{\pi} \mbox{ iff } \varphi=\frac{e^{i\theta}}{\sqrt{2}}\left[\begin{array}{c}1\\-1\end{array}\right]
\]
where $0 \le \theta < 2 \pi.$

Moreover we consider the conditional expectation of the first hitting time to 0 starting from $k=1$ given an event $\{T_0 < \infty \}$, that is, $E_1 ^{(\infty)}(T_0|T_0 < \infty)=E_1 ^{(\infty)} (T_0 ; T_0 < \infty)/P_1 ^{(\infty)} ( T_0 < \infty)$. Let 
\[
f(z)= \sum_{n=1} ^{\infty} (r^{(\infty)} _1 (n)) ^2 z^n 
\left( = {{}_2F_1(-1/2,-1/2;1;z^4)-1 \over z} \right)
\]
In this case we have to know the value of $f'(1)$. It should be noted that
\[
{d \over dz} \left( {}_2F_1(a,b;c; g(z)) \right) 
= \left( {ab \over c} \right) \> {}_2F_1(a+1,b+1;c+1; g(z)) g'(z) 
\]
The above formula gives
\[
f'(z)= { z^4 {}_2F_1(1/2,1/2;2;z^4) - {}_2F_1(-1/2,-1/2;1;z^4)+1 \over z^2}
\] 
Moreover noting   
\[
{}_2F_1(-1/2,-1/2;1;1) = {}_2F_1(1/2,1/2;2;1) = {4 \over \pi} 
\] 
we have $f'(1)=1$. Therefore the desired conclusion is obtained: 
\[ 
  E_1 ^{(\infty)} (T_0|T_0 < \infty)={ \sum_{n=1} ^{\infty} n P_1 ^{(\infty)}(T_0=n) \over \sum_{n=1} ^{\infty} P_1 ^{(\infty)} (T_0 = n) }
  = {1 \over P_1^{(\infty)}(\varphi)}
\]
since
\begin{eqnarray*} 
&& \sum_{n=1} ^{\infty} n P^{(\infty)} _1 (T_0 = n ) \\
&& = 
\sum_{n=1} ^{\infty} n 
\Bigg[ 
{1 \over 2} \left\{ (p^{(\infty)} _1 (n)) ^2 + (r^{(\infty)} _1 (n)) ^2 
\right\}
+{1 \over 2} \left\{ (p^{(\infty)} _1 (n)) ^2 - (r^{(\infty)} _1 (n)) ^2 \right\} (\alpha{\bar\beta}+{\bar\alpha}\beta) 
\Bigg] \\
&& = {1 \over 2} \{ 1 + f'(1) \} + {1 \over 2} \{ 1 - f'(1) \}  
(\alpha{\bar\beta}+{\bar\alpha}\beta) =1
\end{eqnarray*} 
In a similar way we see that $f''(1)= \infty$ implies 
\[
E_1 ^{(\infty)} ((T_0)^2 |T_0 < \infty) = \infty
\]
so the $m$th moment $E_1 ^{(\infty)} ((T_0)^m |T_0 < \infty)$ diverges for $m \ge 2.$

Next we consider the finite $N$ case. Then $p^{(N)} _k (z)$ and $r^{(N)} _k (z)$ satisfy 
\begin{eqnarray*}
   && p^{(N)}_k(z) = A_z \lambda_+^{k-1}+B_z \lambda_-^{k-1} \\
   && r^{(N)}_k(z) = C_z \lambda_+^{k-N+1}+ D_z \lambda_-^{k-N+1}
\end{eqnarray*}
since $\lambda_+ \lambda_- = -1$. All we have to do is to determine the coefficients $A_z,B_z,C_z,D_z$ by using boundary conditions: $p^{(N)}_1(z)=z$ and $r^{(N)} _{N-1}(z)=0$ come from the definition of $\Xi_k ^{(N)} (n)$. The boundary conditions imply $C_z+D_z=0$ and $A_z+B_z=z$, so we see
\begin{eqnarray}
&& p^{(N)}_k(z) = \left( {z \over 2} +E_z \right) \lambda_+^{k-1}
+ \left( {z \over 2} -E_z \right) \lambda_-^{k-1} \\
&& r^{(N)}_k(z) = C_z (\lambda_+^{k-N+1}-\lambda_-^{k-N+1})
\end{eqnarray}
where $E_z= A_z - z/2=z/2 - B_z$. To obtain $E_z$ and $C_z$, we use $r^{(N)} _1(z)=(p^{(N)} _2(z)-r^{(N)} _2(z))z/\sqrt{2}$ and $r^{(N)} _{N-2}(z)=(p^{(N)} _{N-1}(z)-r^{(N)} _{N-1}(z))z/\sqrt{2}=p^{(N)} _{N-1}(z) z/\sqrt{2}$.
Therefore
\begin{eqnarray*}\label{AC}
&& C_z(\lambda_+ - \lambda_-) = 
     \frac{z}{\sqrt{2}}\left\{ \left( \frac{z}{2}+E_z \right) \lambda_+^{N-2}
     + \left( \frac{z}{2}-E_z \right) \lambda_-^{N-2} \right\} \\
&& C_z(\lambda_+^{N-2}-\lambda_-^{N-2}) = 
     \frac{z}{\sqrt{2}}\left\{ \left( \frac{z}{2}+E_z \right) 
     (-1)^{N-1}\lambda_+ + \left( \frac{z}{2}-E_z \right) (-1)^{N-1}\lambda_- 
     + C_z (\lambda_+^{N-3}-\lambda_-^{N-3})\right\}
\end{eqnarray*}
Solving the above equations gives
\begin{eqnarray} 
&& C_z = {z^2 \over \sqrt{2}} 
(-1)^{N-2}(\lambda_+^{N-3} - \lambda_-^{N-3}) \\
&& \times \left\{ 
(\lambda_+^{N-2} -\lambda_-^{N-2})^2
-{z \over \sqrt{2}} (\lambda_+^{N-2} -\lambda_-^{N-2})
(\lambda_+^{N-3} -\lambda_-^{N-3})
-(-1)^{N-3}(\lambda_+ - \lambda_-)^2 \right\}^{-1} \nonumber \\
&& E_z = - {z \over 2(\lambda_+^{N-2} -\lambda_-^{N-2})}
\bigg[ 2 (-1)^{N-3}(\lambda_+ - \lambda_-) 
(\lambda_+^{N-3} -  \lambda_-^{N-3}) \\
&& \times \left\{ 
(\lambda_+^{N-2} -\lambda_-^{N-2})^2
-{z \over \sqrt{2}} (\lambda_+^{N-2} -\lambda_-^{N-2})
(\lambda_+^{N-3} -\lambda_-^{N-3})
-(-1)^{N-3} (\lambda_+ - \lambda_-)^2 \right\}^{-1} \nonumber \\
&& \qquad \qquad \qquad \qquad \qquad \qquad \qquad \qquad \qquad 
+ \> (\lambda_+^{N-2} + \lambda_-^{N-2}) \bigg] \nonumber 
\end{eqnarray} 
By Lemma 1, we obtain 
\begin{thm}
\label{thm:thm2}
\begin{eqnarray*} 
P^{(N)} _k (\varphi) 
= C_1 |\alpha|^2 + C_2 |\beta|^2 + 2 \Re(C_3 \overline{\alpha} \beta)
\end{eqnarray*} 
where $\varphi = {}^t[\alpha, \beta] \in \Phi$ and 
\begin{eqnarray*}
&& C_1 = {1 \over 2 \pi} \int_0 ^{2\pi} |a p^{(N)} _k (e^{i \theta}) + c r^{(N)} _k (e^{i \theta})|^2 d \theta \\
&& C_2 = {1 \over 2 \pi} \int_0 ^{2\pi} |b p^{(N)} _k (e^{i \theta}) + d r^{(N)} _k (e^{i \theta})|^2 d \theta \\
&& C_3 =  {1 \over 2 \pi} \int_0 ^{2\pi}
\overline{(a p^{(N)} _k (e^{i \theta}) + c r^{(N)} _k (e^{i \theta}))} (b p^{(N)} _k (e^{i \theta}) + d r^{(N)} _k (e^{i \theta}) ) d \theta 
\end{eqnarray*}
with $a=b=c=-d=1/\sqrt{2}$, here $p^{(N)} _k (z)$ and $r^{(N)} _k (z)$ satisfy Eqs. (4.19) and (4.20), and 
$C_z$ and $E_z$ satisfy Eqs. (4.21) and (4.22).
\end{thm}
Therefore to compute $P^{(N)} _k (\varphi)$, we need to obtain both $p^{(N)} _k (z)$ and $r^{(N)} _k (z)$, and calculate the above $C_i (i=1,2,3)$ explicitely.

Here we consider $U=H$ (the Hadamard walk), $\varphi = {}^t [\alpha, \beta]$ and $k=1$. From Theorem 2, noting that $p^{(N)} _1 (z) = z$ for any $N \ge 2$, we have 
\begin{cor}
\label{cor:cor3} For $N \ge 2$, 
\begin{eqnarray*}
P^{(N)} _1 (\varphi) ={1 \over 2} \left( 1 + {1 \over 2 \pi} \int_0 ^{2\pi} |r^{(N)} _1 (e^{i \theta})|^2 d \theta \right) 
( 1 + 2 \Re(\overline{\alpha} \beta))
\end{eqnarray*}
where $r^{(2)} _1 (z) = 0, \> r^{(3)} _1 (z) = z^3/(2 - z^2),$ 
\begin{eqnarray*}
r^{(4)} _1 (z) = {z^3 (1 - z^2) \over 
2- 2 z^2 + z^4 }, 
\quad r^{(5)} _1 (z) = {z^3 (2 - 3 z^2 +2 z^4) \over 
4 -6 z^2 +5 z^4 -2 z^6},
\quad 
 r^{(6)} _1 (z) = {2z^3 (1-z^2)(1-z^2+z^4) \over 
4 -8 z^2 +9 z^4 -6 z^6+2z^8}
\end{eqnarray*}
and in general for $N \ge 4$
\begin{eqnarray*}
r^{(N)} _1 (z)= - 
{ z^2 J_{N-3} (z) J_{N-4} (z) \over \sqrt{2} (J_{N-3} (z))^2 -z J_{N-3} (z) J_{N-4} (z) - \sqrt{2} (-1)^{N-3}} 
\end{eqnarray*}
with 
\[
J_n (z) = \sum_{k=0} ^n \lambda_+ ^k \lambda_- ^{n-k}, \quad \lambda_+ + \lambda_- = \sqrt{2} \left( z - {1 \over z} \right), \quad \lambda_+ \lambda_- = -1
\]
\end{cor}
In particular, when $\varphi = {}^t [0,1] = |R \rangle, \> k=1$ and $N=2, \ldots ,6$ cases, the above corollary gives
\begin{eqnarray*}
&& P^{(2)}_1 ({}^t [0,1] ) = {1 \over 2}, 
\quad P^{(3)}_1 ({}^t [0,1] ) = {2 \over 3}, 
\quad P^{(4)}_1 ({}^t [0,1] ) = {7 \over 10}, \\
&& P^{(5)}_1 ({}^t [0,1] ) = {12 \over 17},
\quad P^{(6)}_1 ({}^t [0,1] ) = {41 \over 58}
\end{eqnarray*}
It is easily checked that the above values $P^{(N)}_1 ({}^t [0,1] ) \> (N=2, \ldots ,6)$ satisfy the conjecture given by Eq. (3.15).

\section{Summary}

In this paper we consider absorption problems for quantum walks on $\{0,1, \ldots N \}$ for both $N< \infty$ and $N= \infty$ cases by using the PQRS method. Here we summarize our main results. 

\par
First we describe $N= \infty$ case. In this case, we have
\begin{eqnarray*}  
P^{(\infty)} _1 (\varphi) = {2 \over \pi} + 
2 \left(1 - {2 \over \pi} \right) \Re(\overline{\alpha} \beta)
\end{eqnarray*} 
and
\begin{eqnarray*}  
&& E_1 ^{(\infty)} (T_0|T_0 < \infty) = {1 \over P_1^{(\infty)}(\varphi)} \\
&& E_1 ^{(\infty)} ((T_0)^m |T_0 < \infty) = \infty  \quad (m \ge 2) 
\end{eqnarray*} 
for any initial qubit state $\varphi = {}^t [\alpha, \beta] \in \Phi$, where $P^{(\infty)} _1 (\varphi)$ is the probability that the particle first hit location 0 starting from location 1 and  $E_1 ^{(\infty)} ((T_0)^m |T_0 < \infty)$ is the conditional $m$th moment of $T_0$ starting from location $1$ given an event $\{T_0 < \infty \}$ with $T_0$ is the first hitting time to location 0.
\par
Next we describe $N < \infty$ case. In this case, we obtain the following explicit expression of $P^{(N)} _1 (\varphi)$: 
\begin{eqnarray*}
P^{(N)} _1 (\varphi) ={1 \over 2} \left( 1 + {1 \over 2 \pi} \int_0 ^{2\pi} |r^{(N)} _1 (e^{i \theta})|^2 d \theta \right) 
( 1 + 2 \Re(\overline{\alpha} \beta))
\end{eqnarray*}
for any initial qubit state $\varphi = {}^t [\alpha, \beta] \in \Phi$, where $P^{(N)} _1 (\varphi)$ is the probability that the particle first hit location 0 starting from location 1 before it arrives at location $N$ and $r^{(N)} _1 (z)$ is given by Corollary 3 in the previous section. 

\par
The above our result guarantees that the conjecture presented by Ambainis {\it et al.} (see Eq. (3.15)) is true for $N=2, \ldots , 6$. However, their conjecture is still open for any $N$. So one of the future interseting problems is to find and prove an explicit form like Eq. (3.15) for any $N (\le \infty), \> k \in \{1, \ldots, N-1 \}$, and $\varphi \in \Phi.$ 

\par
\
\par\noindent
{\bf Acknowledgments.}  This work is partially financed by the Grant-in-Aid for Scientific Research (B) (No.12440024) of Japan Society of the Promotion of Science. 

\par
\
\par\noindent

\begin{small}

\bibliographystyle{plain}

\par
\vskip 1.0cm

Department of Applied Mathematics

Faculty of Engineering

Yokohama National University

Hodogaya-ku, Yokohama 240-8501, Japan

norio@mathlab.sci.ynu.ac.jp

\par
\vskip 1.0cm

Division of Mathematics

Graduate School of Science

Hokkaido University

Kita-ku, Sapporo 060-0810, Japan

nami@math.sci.hokudai.ac.jp

\vskip 1cm

Department of Mathematics and Statistics

Monash University

Clayton, Victoria 3168, Australia 

Aidan.Sudbury@sci.monash.edu.au\\

\vskip 1cm

\end{small}

\end{document}